\documentstyle[a4]{article}
\begin{document}                
\newcommand{\manual}{rm}        
\newcommand\bs{\char '134 }     

\newcommand{\simlt}{\stackrel{<}{{}_\sim}}
\newcommand{\simgt}{\stackrel{>}{{}_\sim}}
\newcommand{\MeV}{\;\mathrm{MeV}}
\newcommand{\TeV}{\;\mathrm{TeV}}
\newcommand{\GeV}{\;\mathrm{GeV}}
\newcommand{\eV}{\;\mathrm{eV}}
\newcommand{\cm}{\;\mathrm{cm}}
\newcommand{\s}{\;\mathrm{s}}
\newcommand{\sr}{\;\mathrm{sr}}
\newcommand{\lab}{\mathrm{lab}}
\newcommand{\e}{\mathrm{e}}
\newcommand{\ts}{\textstyle}
\newcommand{\ol}{\overline}
\newcommand{\be}{\begin{equation}}
\newcommand{\ee}{\end{equation}}
\newcommand{\ba}{\begin{eqnarray}}
\newcommand{\ea}{\end{eqnarray}}
\newcommand{\nn}{\nonumber}
\newcommand{\nm}{{\nu_\mu}}
\newcommand{\pp}{$\overline{p}(p)-p\;\;$}
\renewcommand{\floatpagefraction}{1.}
\renewcommand{\topfraction}{1.}
\renewcommand{\bottomfraction}{1.}
\renewcommand{\textfraction}{0.}
\renewcommand{\thefootnote}{F\arabic{footnote}}
\title{The possibility of a sizable, direct CP-violating asymmetry in
$B^\mp \to K^\mp \eta$}
\author{Saul Barshay and Georg Kreyerhoff\\III. Physikalisches Institut\\
RWTH Aachen\\D-52056 Aachen\\Germany} \maketitle
\begin{abstract}                
The likelihood of a sizable, direct CP-violating asymmetry in the
decays $B^\mp\to K^\mp\eta$, is calculated within the framework
of the model which originally predicted a sizeable asymmetry in
$\pi^\mp\eta(\eta')$. It is shown in a transparent manner that
these decays are the best places for providing the first clear
evidence for direct CP violation in the decays of a charged
particle. These decays contain two amplitudes of comparable
magnitude, with different weak-interaction phase, and with
different strong-interaction phase which is explicitly calculated
in a model of three coupled channels of physical hadrons.
\end{abstract}

Recently, the BABAR collaboration reported the observation of the
decays \newline $B^\mp\to \pi^\mp\eta$, with a small branching
ratio of $(4.2\pm 1.0 \pm 0.3)\times 10^{-6})$ \cite{ref1}. It is
notable that there is an indication of a sizable, direct
CP-violating asymmetry\newline ${\cal A}(B^- - B^+)= (-0.51\pm
0.19 \pm 0.01)$ \cite{ref1}. At present, this is potentially the
first distinctly nonzero asymmetry ever detected in the decay of
a charged particle. That this particular decay mode with a small
branching ratio provides the best possibility for observing
direct CP violation in charged-meson decays, was predicted in
1991 \cite{ref2}. The prediction is based upon detailed
calculations of partial rates and asymmetries, which utilize
known, strongly-coupled channels of physical hadrons
$\pi^\mp(\eta,\eta',\eta_c)$ in order to calculate the
final-state interaction phases which are essential for obtaining
a direct, CP-violating asymmetry \cite{ref2,ref3}. The other
necessary ingredient \cite{ref2,ref3} is the presence of two
weak, effective amplitudes, each with a different CP-violating
phase (given as functions of the phase $\delta_{13}$ of the CKM
quark-mixing matrix). The CP-violating asymmetry arises from the
interference between these two amplitudes; it can be sizable when
the amplitudes have comparable magnitude \cite{ref2,ref3}. This
is the case for $B^\mp \to \pi^\mp \eta (\eta')$, because one
tree-level diagram \footnote{The $s_{ij}$ denote sines of
quark-mixing angles in the usual form \cite{ref4} of the CKM
matrix, with $\delta_{13}$ the large, CP-violating phase;
$\alpha_s\sim 0.2$ is the QCD coupling at a scale $\sim
m_{B^0}\sim 5 \GeV$ \cite{ref4}. Numerical values of the $s_{ij}$
and $\delta_{13}$ are taken from \cite{ref4}.} for $b\to u + (W\to
\ol{u}d)$ is proportional to $|s_{13} \e^{-i \delta_{13}}| \sim |
(s_{12}s_{23}/2.5)\e^{-i\delta_{13}}| $; and the second tree-level
diagram for $b\to c + (W\to \ol{c}d)$ is proportional to
$|s_{23}s_{12}|$. From the strong-interaction transition $\eta_c
\pi \to \eta (\eta') \pi$, the second tree diagram acquires a
multiplication factor of $+i$ times an effective phase angle,
approximately calculated \cite{ref2} to be a number of the order
of $0.03$.\footnote{In the original paper \cite{ref2}, the
$K$-matrix elements are underestimated by a factor of 2.
Incorporating this factor, increases the two direct, CP-violating
asymmetries for $\eta,\eta'$ by a factor of 2. (The CP-violating
phase denoted by $\phi$ is $\delta_{13}$.) } From interference
with the first tree diagram, one immediately obtains an order of
magnitude estimate for the direct, CP-violating asymmetry, ${\cal
A} \sim -\frac{2(0.03)(1/2.5)}{(1/2.5)^2}\sin \delta_{13} \sim
-0.15 \sin \delta_{13}$. For $B^\mp \to \pi^\mp \eta(\eta')$, the
loop diagrams involving $b\to t,c$ and $t,c \to d$, with gluon
emission and conversion into a $\ol{u}u$ quark pair are
proportional to $|s_{12} s_{23}| \times \alpha_s/\pi$, and so are
down by a factor of the order of$^{F1}$ $\alpha_s/\pi \sim 0.07$.
The small branching ratio for $B^\mp \to \pi^\mp \eta'$ has very
recently been reported \cite{ref5}, $(2.8^{+1.3}_{-1.0} \pm
0.6)\times 10^{-6}$. It appears to be comparable to, but a little
less than the branching ratio for $\pi^\mp \eta$, as predicted
\cite{ref2}. A sizable CP-violating asymmetry is expected
\cite{ref2}.

In this note, we discuss three positive and one negative result
which follow directly from the model which led to the above
predictions, when the model is extended to the decays $B^\mp \to
K^\mp (\eta,\eta',\eta_c)$. Within the context of the quark
diagrams considered in the model, we discuss two possible
remedies for the negative result which concerns the quite large
empirical branching ratio for the decays $B^\mp \to K^\mp \eta'$.
We then consider the explanation in terms of additional quark
diagrams, namely the weak loop (``penguin'') diagrams, involving
$b\to t,c$ and $t,c\to s$, with gluon emission and conversion
into an $\ol{s}s, \ol{u}u$ pair. Such diagrams are generally
considered to have an essential role \cite{ref6} in the decays
$B\to \pi K$, and also in $B^\mp \to K^\mp \eta(\eta')$
\cite{ref7}. Explanations of the large $\eta'$ branching ratio in
the model without additional loop diagrams are possible, but do
not seem to be adequate. Contributions from both dynamics are
probably necessary. A new contribution involving a small
$\ol{c}c$ component in $\eta'$ is considered, and can tested for
experimentally \cite{ref3}. Our main positive results concern the
empirically small branching ratio for $B^\mp \to K^\mp \eta$, and
most importantly for the experimental situation \cite{ref1}, the
possibility of a sizable, direct CP-violating asymmetry ${\cal A}
\sim -0.2 \sin \delta_{13}$, for this decay mode. The recent
report \cite{ref1} which indicated a sizable asymmetry for
$\pi^\mp\eta$, also gave an asymmetry for $K^\mp\eta$, of
$-0.32\pm 0.20 \pm 0.01$. This is a principal motivation for the
present analysis, because this could be a second place  where
sizeable, direct CP violation can be anticipated and observed in
the decay of a charged meson. It is valuable for experimentalists
to have a transparent estimate for direct CP violation in
charged-meson decay for those few modes where a sizable asymmetry
can likely be clearly observed \cite{ref1}. Such clear
observations would be a milestone in the history of CP
noninvarinance \cite{ref8,ref9,ref10}.

The first result follows immediately from the number obtained in
1991 for the branching ratio for $B^\mp \to \pi^\mp \eta_c$,
given as about $4.1 \times 10^{-5}$ for the decay constant
$f_{\eta_c} \sim 310 \MeV$. From this number, the branching ratio
for $B^\mp \to K^\mp \eta_c$ is obtained approximately by simply
dividing by $s_{12}^2 \cong 0.48$; it is thus predicted to be
quite large, $\sim 1\times 10^{-3}$ ( for $f_{\eta_c} \cong 335
\MeV$ \cite{ref11} ). Recent experimental results confirm this:
CLEO \cite{ref11} gives $(0.69^{+0.26}_{-0.21} \pm 0.08 \pm 0.20
) \times 10^{-3}$; BELLE \cite{ref12} gives $(1.25 \pm 0.14
^{+0.10}_{-0.12} \pm 0.38 ) \times 10^{-3}$. The first error is
statistical, the second is systematic, and the third is from the
$\eta_c$ branching-fraction uncertainty. The large branching
ratio supports the essential validity of the phenomenological,
effective weak hamiltonian \cite{ref13}, and of the physical
hadron ``factorization'' approximation \cite{ref13,ref14,ref15}
used in calculating matrix elements for particular exclusive
decay modes (that is, matrix elements of products of weak
currents factorize into matrix elements of individual currents in
states of physical hadrons). We explain next the second and third
results, concerning the small branching ratio for $B^\mp \to
K^\mp \eta$, and concerning the distinct possibility of a
sizable, direct CP-violating asymmetry. From eq. (27) in
\cite{ref2}, the form of the full, ``unitarized'' \cite{ref16}
amplitude (denoted by $A^u$) \footnote{The three amplitudes
$A_{\eta,\eta',\eta_c}^u$ satisfy CPT invariance \cite{ref2},
namely that the widths of $B^-$ and $B^+$, summed over all three
channels $K^\mp(\eta,\eta',\eta_c)$ are equal. } is$^{F2}$ \ba
A_\eta^u
&\cong& A_\eta + i t_{12} A_{\eta'} + i t_{13} A_{\eta_c}\nn\\
 & \sim & A_\eta - i (0.026) A_{\eta_c}\ea
The $A$ are the weak amplitudes following from the effective,
weak hamiltonian, computed with the factorization approximation.
The $t_{ij}$ ($i,j = 1,2,3 = \eta,\eta',\eta_c)$ are given
approximately \cite{ref2} by linear combinations of the tangents
of the three small eigenphases; these were originally calculated
\cite{ref2} for the specific system of three coupled channels
$\pi^\mp (\eta,\eta'\eta_c)$. The essential number is$^{F2}$
$|t_{13}|\sim 0.026$ calculated in \cite{ref2}; we assume that
this has about the same magnitude for the system of three coupled
channels, $K^\mp(\eta,\eta',\eta_c)$.\footnote{The experimental
limit \cite{ref4} for the branching ratio for $\eta_c \to
K\ol{K}\eta$ is $<3.1\%$, comparable to the measured $(4.9\pm
1.8)\%$ for $\pi\pi\eta$ and $(4.1\pm 1.7)\%$ for $\pi\pi\eta'$.
(That for $K\ol{K}\pi$ is $(5.5\pm 1.7)\%$.) The $t_{13},t_{23}$
are related to the square root of the branching ratios
\cite{ref2}. Only if $\eta_c\to K\ol{K}\eta(\eta')$ were to be
very small would they be much smaller than used here, and hence
the asymmetry in $K^\mp\eta$ smaller. } There is a sign ambiguity
related to the analogue for $K^\mp$ of eq.~(18) in \cite{ref2} for
$\pi^\mp$. Here we choose the negative sign, which leads to a
negative $t_{13}$. Except for the CKM factors, the magnitude of
the weak amplitudes are taken as approximately the same as
calculated in \cite{ref2}.\footnote{Using the approximation of
the same decay constant and wave-function overlap for $K^\mp$ as
for $\pi^\mp$, as given in \cite{ref2} ($f_{\pi^+}\sim 130\MeV$
\cite{ref4}). } In eq.~(1), $A_{\eta_c}$ is proportional to
$|s_{12}s_{13} \e^{-i\delta_{13}}| \sim \left|\frac{s_{12}^2
s_{23}}{2.5}\e^{-i\delta_{13}}\right| \sim (0.02)s_{23}$. In this
initial discussion, we neglect the term in $A_{\eta'}$ which, in
the present approximation, is similar in magnitude \cite{ref2} to
$A_\eta$, and is multiplied by the calculated \cite{ref2} small
number $|t_{12}|\sim 0.1$.$^{F2}$ Below, we discuss the addition
of the loop diagram that can give a large $A_{\eta'}$ in eq.~(1),
and which can then make this term comparable to the first term
(which is also enhanced somewhat). Our main result here,
concerning the possibility of a sizable CP-violating asymmetry in
$B^\mp\to K^\mp\eta$, will not be changed by inclusion of an
enhanced $A_{\eta'}$. The branching ratio for $K^\mp\eta$ is
increased, and the asymmetry correspondingly reduced. Computing
the explicit numerical value \cite{ref2} of $A_\eta$, and
expressing it as a multiple of the product of the amplitude
arising from $A_{\eta_c}$ and the strong final-state interaction
factor, we obtain$^{F2}$ \be A_\eta^u \sim -\left\{
\left(\frac{17}{39}\right) \e^{-i\delta_{13}}+i\right\} (0.026)
A_{\eta_c}\ee Thus, there is a clear demonstration of two
amplitudes of comparable magnitude, differing in weak-interaction
phase and in strong-interaction phase. With the large branching
ratio for $K^\mp\eta_c$ of about $10^{-3}$, this gives the small
branching ratio for $K^\mp\eta$ of about \ba &&
\frac{1}{2}\left\{ \mathrm{b.r.} (B^- \to
K^-\eta)+\mathrm{b.r.}(B^+\to K^+\eta)\right\}\nn\\ &\sim&
\left\{1+(17/39)^2\right\}\times(0.026)^2\times 10^{-3}\cong 0.8
\times 10^{-6}\ea The model gives a quite small branching ratio
for $K^\mp\eta$, essentially in terms of the large branching
ratio for $K^\mp\eta_c$ and a calculated, small final-state
interaction factor. Recent measurement \cite{ref1} gives for the
branching ratio for $K^\mp\eta$, $(2.8\pm 0.8\pm 0.2)\times
10^{-6}$. From eq.~(2) the CP-violating asymmetry is calculated
as \be {\cal A} (B^- - B^+) \cong
-\frac{2(17/39)\sin\delta_{13}}{\left\{1+(17/39)^2\right\}}\sim
-0.73\sin\delta_{13}\ee The possibility of an asymmetry which is
quite sizable in magnitude occurs naturally in this model.

The negative result within the model is a small branching ratio
for $B^\mp \to K^\mp\eta'$. The analogue of eqs.~(1,2) is
\cite{ref2}$^{F2}$ \ba A_{\eta'}^u &\cong& A_{\eta'} + i t_{21}
A_\eta + i t_{23} A_{\eta_c} \cong A_{\eta'} -i (0.04)
A_{\eta_c}\nn\\
&\sim& - \left\{ \left(\frac{8}{39}\right)
\e^{-i\delta_{13}}+i\right\} (0.04) A_{\eta_c}\nn\\&\Rightarrow&
\frac{1}{2}\left\{ \mathrm{b.r.}(B^- \to K^-\eta') +
\mathrm{b.r.}(B^+\to K^+\eta')\right\} \sim 1.7\times 10^{-6}\ea
The experimental numbers are about 40 times larger, $(6.5\pm
1.5\pm 0.9)\times 10^{-5}$ and $(80\pm 10\pm 7 9\times 10^{-6}$
from CLEO \cite{ref17}, $(79\pm 12\pm 9) \times 10^{-6}$ from
BELLE \cite{ref18}, and $(70\pm 8\pm 5)\times 10^{-6}$ from BABAR
\cite{ref19}. So a factor of about 6.3 larger in amplitude.
Within the context of the quark diagrams considered so far in the
model, there are two ways to increase the amplitude.
\begin{itemize}
\item[(1)] Constructive contributions from additional (to
$\eta_c$) $\ol{c}c$ states could contribute to enhancing an
effective $|t_{23}|$. Specifically, the state $\chi_{c1}$ and the
higher-mass $\eta_c$ state(s) may have significant decay
amplitudes to $\eta' K^+ K^-$. It is notable that $\chi_{c1}$
does have significant, measured \cite{ref4} branching ratios to
$\pi^- K^+ K^0_S$ and to  $\pi^+ \pi^- K^+ K^-$. However,
invoking a hypothetical factor of 3 increase in an effective
$|t_{23}|$, to $\sim 0.12$, still leaves the branching ratio of
$\sim 14\times 10^{-6}$ below the experimental value.
\item[(2)] There could be a small $\ol{c}c$ component in $\eta'$.
Applied to $\eta$, this is an old idea due to Feldman and Perl
\cite{ref20}, who considered it as a possible explanation for the
relatively large branching ratio for $\Psi' \to J/\Psi\; \eta$.
Such a component, at the level of $<15\%$ in amplitude, was used
in \cite{ref3} to calculate two weak amplitudes of comparable
magnitude, whose interference led to sizable, direct CP-violating
asymmetries in the angular distributions and rates for
$B^0(\ol{B})^0\to \pi^+\pi^- \eta$. However, invoking a
hypothetical $\ol{c}c$ amplitude in $\eta'$ of about $15\%$, again
raises the $K^\mp\eta'$ branching ratio to only about $23\times
10^{-6}$. A direct experimental test for such an amplitude would
be in the then expected \cite{ref3} sizable, direct CP violation
in angular distributions and rates for $B^0(\ol{B}^0)\to \pi^-
\pi^+ \eta'$.
\end{itemize}

We turn to the weak loop diagrams involving $b\to t,c$ and
$t,c\to s$, which are proportional to $s_{23}(\alpha_s/\pi)$ and
which seem to be necessary to explain the large branching ratio
for $K^\mp \eta'$. Our main concern here is the implication of an
enhanced $\eta'$ amplitude for the direct, CP-violating asymmetry
estimated above in eq.~(4) for the small-branching mode
$K^\mp\eta$. Lipkin noted \cite{ref7} that the negative sign of
the $\ol{s}s$ amplitude in the $\eta$, as compared to the
positive sign in the $\eta'$, simply leads to a strong
suppression of $B^\mp \to K^\mp\eta$, when the decay amplitude
involves the $b\to s+(g\to \ol{s}s)$ loop diagram with gluon
emission and conversion into an $\ol{s}s$ pair. Indeed, if this
decay amplitude dominates and is responsible for an experimental
\cite{ref17,ref18,ref19} b.r.$(K^\mp\eta')\sim 70\times 10^{-6}$,
then a possible suppression factor \cite{ref7} of the order of
$1/34$ brings one to $1/2\{ \mathrm{b.r.}(K^-\eta) +
\mathrm{b.r.}(K^+\eta)\}$ of $\sim 2\times 10^{-6}$, close to the
experimental \cite{ref1} number. This squared amplitude would
increase the numerical value of the denominator in eq.~(4) by a
factor of the order of 3. If this amplitude contributes little to
the numerator (i.~e.~if it has negligible phase factors), then
the asymmetry in eq.~(4) would be reduced to ${\cal A} \sim
-0.24\sin \delta_{13}$. With an enhanced weak amplitude
$A_{\eta'}$, the term $it_{12}A_{\eta'}$ in eq.~(1) also
contributes to $A_\eta^u$ (with $|t_{12}|\sim 0.1)$$^{F2}$. This
term could increase the numerator in eq.~(4), as well as the
denominator; including it constructively, we estimate ${\cal
A}\sim -0.2\sin \delta_{13}$. It is an open question \cite{ref17}
as to whether the loop diagrams alone can actually account for
the $\eta'$ branching ratio. Assuming $b\to s + g$ to have a
branching fraction of about $0.5\%$ \footnote{Note that reducing a
branching ratio of $\sim 5\times 10^{-3}$ by $\alpha/\alpha_s\sim
(1/137)/0.2 \cong 1/27$ results in $\sim 1.8\times 10^{-4}$, only
just below the experimental branching ratio \cite{ref4} of
$(3.1\pm 1.1)\times 10^{-4}$ for $b\to s+\gamma$. A dominant
exclusive channel $K^* (892)\gamma$, is down by $1/8$
experimentally \cite{ref4}.}, then a (quasi) exclusive mode, like
$(K^\mp\pi^0 + K^0\pi^\mp)$, if largely ``penguin'' mediated,
would have a branching ratio of the order of
$(1/10)(\alpha_s/\pi)(5\times 10^{-3})\sim 3.5\times 10^{-5}$,
\newline with $\sim \alpha_s/\pi\sim 0.07$ arising from the gluon
emission and conversion into $\ol{u}u$, $\ol{d}d$, and the factor
of $\sim 1/10$ approximately accounting for a (quasi)exclusive
mode.$^{F6}$ This simple estimate is remarkably close to
experiment \cite{ref4} for $K\pi$, but is still somewhat below
$K^\mp\eta'$ \cite{ref16}. In addition, there is a small
contribution from the amplitudes giving rise to eq.~(5). In this
situation, a diagram involving a hypothetical, small (up to $\sim
18\%$ in amplitude) $\ol{c}c$ component \cite{ref20,ref3} in
$\eta'$, if constructive with the loop diagram, could increase
the branching ratio to experiment. As mentioned above, a direct
experimental test for such a component is possible in the decays
$B^0(\ol{B}^0)\to \pi^- \pi^+ \eta'$. The effect upon the
$K^\mp\eta$ asymmetry is to increase it somewhat $(\sim 1.4)$,
because of a decrease in the calculated $K^\mp\eta$ partial rate
from the smaller loop contribution. In contrast to the prediction
of a sizable, direct CP-violating asymmetry in $K^\mp\eta$, the
prediction for $K^\mp\eta'$ is for a small asymmetry, of the
order of a few percent. This is because of the dominant size of
the loop diagram (approximately independent of $\delta_{13}$), in
determining the branching ratio. An interference term normalized
to the partial rates, giving rise to an asymmetry, must be down
by a factor of at least of the order of $1/10$ from eq.~(4).
Interestingly, the experimental number \cite{ref4}, appears to be
small, ${\cal A}_{K^\mp\eta'}=-0.02\pm 0.07$.

In summary, in addition to the original prediction \cite{ref2}
for $B^\mp\to \pi^\mp\eta(\eta')$, the extended model predicts
the likelihood of a sizable, direct CP-violating asymmetry in
$B^\mp\to K^\mp\eta$, but not in $K^\mp\eta'$. The physical
origin of these predictions is the presence in the three decay
processes of two amplitudes which are comparable in magnitude,
and which contain different CP-violating weak phases and
different strong-interaction phases \cite{ref3}. The small
comparable magnitudes result from small relevant factors in the
CKM matrix {\underline{and}} from the small magnitude of
(imaginary) numbers which are obtained from calculating
\cite{ref2} the strong-interaction phases in a specific model of
three coupled channels of physical hadrons.\footnote{The use of
coupled channels for calculating strong-interaction phases has
been applied in trying to understand recent data on CP violation
in $B^0(\ol{B}^0)\to \pi^+\pi^-$. S.~Barshay and G.~Kreyerhoff,
hep-ph/0303254v4; JHEP in press.} Decay modes with relatively
large branching ratios, like $K^\mp\eta'$, tend to be dominated
by a single, large decay amplitude, and hence have small
asymmetries. Thus, $B^\mp\to \pi^\mp \eta(\eta')$ and
likely$^{F4}$ $K^\mp\eta$, are the best processes in which direct
CP violation can be established for the first time in the decay
of a charged particle.

S.~B.~thanks Lahlit Sehgal for information and constructive
criticism.

\end{document}